\documentclass{PoS}

\title{Interaction between vortices in CFL phase}

\ShortTitle{Interaction between Vortices in CFL Phase}

\author{\speaker{Sedigheh Deldar}\\
        
        University of Tehran\\
        E-mail: \email{sdeldar@ut.ac.ir}}

\author{Hadi Lookzadeh\\
       University of Tehran\\
        E-mail: \email{h.lookzadeh@ut.ac.ir}}

\abstract{We try to calculate the interaction between vortices  in the color-flavor locked (CFL) phase of Quantum Chromodynamics (QCD). The
 most fundamental strings in high density color superconductivity are the non-abelian semi-super fluid strings. Using Abrikosov ansatz, one can 
show that the interaction between these vortices has a universal repulsion form when they are far from each other. The Ginzburg-Landau (GL) 
Lagrangian of CFL phase may be compared with the Lagrangian of Quantum Electrodynamics (QED) with three condensation states. The behavior 
of vortices of multi-component superconductivity, within the framework of GL theory  was done for two condensation 
states ($\rm{MgB_2}$ superconductors) by variational method. The interaction between these vortices is attractive when they are far from each other and repulsive 
when they get close. We attempt to obtain the interaction between these vortices in such substances using numerical variational method for three
  condensation states (Fe-based superconductors). The ultimate aim of the research is to apply these calculations to the vortices in the color-flavor 
locked phase of QCD to be able to study the interaction between them.}

\FullConference{Xth Quark Confinement and the Hadron Spectrum,\\
		October 8-12, 2012\\
		TUM Campus Garching, Munich, Germany}

\begin{document}

\begin {figure}
\begin{center}
 {\includegraphics[height=0.2\textheight, width=0.9\textwidth]{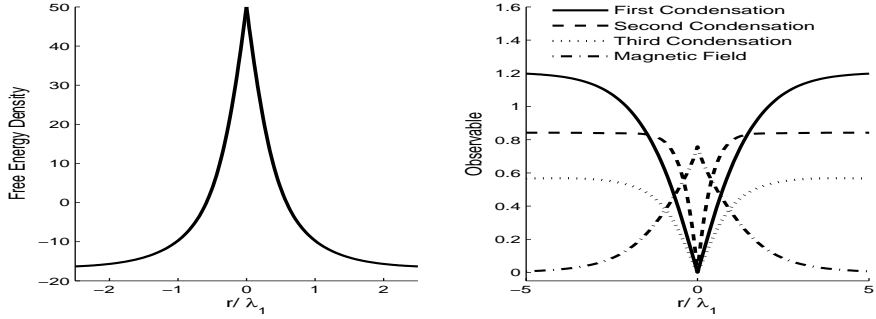}}
\end{center}
\vspace{-20pt}
\caption{\label{f1} Left: Free energy density for a vortex with three condensation states.
 Right: Three condensation states and magnetic field versus distance from the center of the vortex.}
\end {figure}

Quark-quark pairing breaks the symmetry of QCD to CFL phase symmetry
$(G=SU(N)_{\rm C} \times SU(N)_{\rm F}\ \times U(1)_{\rm B}\to H=SU(N)_{\rm F+C}\times Z_N) $.
Since the remaining symmetry is not abelain and the first fundamental group is not trivial, non-abelian vortices emerge \cite{non-abelian}
$(\space G \to H,
\space \space \pi _1(G/H)\neq I,
\pi_1(G/H\thickapprox U(N))=Z)$.
We are interested in the interaction between these non-abelian vortices.   Abrikosov ansatz can be used when they are far enough. But the question is how these non-abelian vortices 
interact when they are close. By expanding the Lagrangian of the non-abelian vortices one can see that it is like the QED Ginzburg-Landau Lagrangian for multi-bands 
superconductors with a Josephson coupling.  We  plan to use the method of calculating the interaction between vortices introduced by Rebbi\cite{Rebbi}.
Generalizing the G-L Lagrangian to  the CFL phase of QCD, one gets
\begin{eqnarray}\label{GL}
\vspace{-1 pt}
{\cal L} =
 {\rm{tr}}({D}\Phi)^{\dag}({D}\Phi)
-m^2{\rm{tr}} (\Phi^{\dag}\Phi)
-\lambda_1 ( {\rm{tr}}\Phi^{\dag}\Phi)^2
-\lambda_2 {\rm{tr}}\left[(\Phi^{\dag}\Phi)^2\right] 
-\frac{1}{4}F^a_{ij}F^{a   ij},
\vspace{-1 pt}
\end{eqnarray}
where $ \Phi$ is the matrix field corresponds to the pairing gap which its elements are defined by
\begin{eqnarray}\label{Phi}
 \Phi_{\alpha i} \sim  \epsilon_{\alpha\beta\gamma}
\epsilon_{ijk} \langle \psi^{T\beta}_{j} C \gamma_5 \psi^{\gamma}_{k} \rangle.
\end{eqnarray}
 These are the color and flavor antisymmetric representations of diquark Cooper pairs with spin zero. 
$\alpha$, $\beta$, and $\it{etc}$  are flavor indices and $i$, $j$, and $\it{etc}$  are color indices. $C$ is the charge conjugation operator and $\psi$ indicates the quark field.
The fundamental string, non-abelian string, is generated by both $SU(N)_{C,F}$ and the $U(1)_B$ generators.
A string with a symmetry along the z-axis in polar coordinate $(\rho,\theta)$ can be shown by\\
\begin{eqnarray}\label{string}
\begin{array}{l}
 \Phi (\theta, \rho) =
\exp \left(i \frac{\theta}{N}\right) 
\exp \left(-iT_{N^2-1} \frac{\sqrt{N(N-1)}}{N}\theta \right) 
{\rm diag}\left(f(\rho), g(\rho), \cdots , g(\rho) \right)= {\rm diag}(e^{i\theta} f, g, \cdots , g)
\end{array}
\end{eqnarray}
The interaction between vortices when they are close to each other has already been studied in QED\cite{Rebbi,con}. 
There exist two  types of superconductors \cite{con} defined by the G-L parameter. It is type I if $k=\lambda/\xi < 1/\sqrt{2}$ .  $\lambda$ is the London penetration depth and $\xi$ 
is the correlation length. It is type II if $k=\lambda/\xi > 1/\sqrt{2}$.
Possible interactions between QED vortices are
attraction for superconductor type I and repulsion for  superconductor type II.
There is a possibility of attraction when the vortices are far from each other and a repulsion when they get close. This type of superconductor is 
called type 1.5, which is somehow new to the physics world.
 In these substances we have two condensation states. Studying the behavior of substances with three condensations is interesting.  We hope to use 
the results of this type of superconductors in the 
CFL phase of QCD.
In this research,  we use the variational method to study the interaction between vortices. First, we apply this method to the real superconductors.
 The Lagrangian  of a
substances with three condensations is
 \begin{equation}\label{eqs1n3}
\begin{array}{l}
\mathcal{L}=\sum _{i=1,2,3}\left[\frac{\alpha _i}{\left|\alpha _1\right|}\left|\Psi _i\right|{}^2+\frac{\beta _i}{2\beta _1}\left|\Psi _i\right|{}^4+\frac{m_1}{m_i}\left|\left(\frac{1}{i \kappa_1 }\nabla -\textbf{A}\right)\Psi _i\right|^2\right]\\
+(\nabla \times \textbf{A})^2 -\gamma_1 (\Psi _1^*\Psi _2+\Psi _2^*\Psi _1) -\gamma_2 (\Psi _2^*\Psi _3+\Psi _3^*\Psi _2) -\gamma_3 (\Psi _1^*\Psi _3+\Psi _3^*\Psi _1),
 \end{array}
\end{equation} 
All coefficients and symbols are defined in \cite{Tinkham}. For simplicity we consider the case when all couplings between condensations are equal ($\gamma=\gamma_i$).
We look for the solution 
$\Psi _i=f_i(r)e^{i n \theta }$ and $\textbf{A}=\frac{n a(r)}{\kappa _1 r}\textbf{e}_{\theta }$.
The interaction will be realized if we know the solution of these equations for all range of distances. 
The following trial function is used\\
\vspace{-20pt}
\begin{equation}\label{eqs2n1}
\begin{array}{l}
f_1(r)=\sqrt{1+\gamma  \eta }+\exp \left(-\frac{r}{\sqrt{2}\xi_v}\right)\sum _{l=0}^n\left(f_{1,l}\left.r^l\right/l!\right),\\
f_{i=2,3}(r)=\sqrt{\frac{\beta _1}{\beta _i}\left(\frac{\alpha _i}{\alpha _1}+\frac{\gamma }{\eta }\right)}+\exp \left(-\frac{r}{\sqrt{2}\xi_v}\right)\sum _{l=0}^n\left(f_{i,l}\left.r^l\right/l!\right),\\
a(r)=1+\exp \left(-\frac{r}{\lambda_v}\right)\sum _{l=0}^n\left(a_l\left.r^l\right/l!\right) ,
\end{array}
\end{equation}

where $ f_{1,l} $, $ f_{2,l} $, $ f_{3,l} $  and $a_l$ are the variational parameters.
 $\eta$ is introduced in \cite{con}. Using the Newton method with the iteration procedure
$V_i^{(m+1)}=V_i^{(m)}-\sum _j\left[\mathbf{H}^{-1}\right]{}_{ij}D_j^{(m)}$, the variational parameters are obtained. The superscript $m$ represents the value at the $m$th step.
$D_i=\partial \mathcal{F}\left/\partial _{V_i}\right.|_{V_i=V_i^{(m)}}$,
and the Hessian matrix is $H_{ij}=\partial ^2\mathcal{F}/\partial V_i\partial V_j|_{V_{i,j}=V_{i,j}^{(m)}}$.
The stationary solution to the iteration procedure corresponds to the minimum of the free energy.
 Figure \ref{f1} shows the free energy density versus the distance from the center of a vortex. The calculated free energy is finite, as expected. The plot on the right hand side 
of figure 1 shows the condensation states and the magnetic field of a vortex with distance. We choose $\xi_1=51nm$, $ \lambda_1=25nm$; $\xi_2=8nm$, $\lambda_2=30nm$ and $\xi_3=25nm$, $\lambda_3=51nm$
 so that the first condensation is of type I and 
the two others are of type II. 
 At asymptotic distances from the center of the vortex, magnetic field vanishes and each condensation obtains its final value, depending 
on its correlation length and penetration depth. In the next step, we will put two vortices at different distances to calculate the interaction between 
them. Eventually, we would like to apply this method to the vortices of the CFL phase of QCD and study their interactions.\\

\end{document}